\documentclass[aps,twocolumn,pra,superscriptaddress,nofootinbib,10pt]{revtex4-1}
\usepackage[]{amsmath}
\usepackage{graphics}
\usepackage{amssymb}
\usepackage{graphicx}
\usepackage{color}

\newcommand{\be}{\begin{equation}}
\newcommand{\ee}[1]{\label{#1} \end{equation}}

\newcommand {\ket}[1]{\lvert \, #1\rangle}
\newcommand {\bra}[1]{\langle #1 \, \rvert}
\newcommand {\braket}[2]{\langle #1 \, | \, #2 \rangle}

\def\Tr{ {\textrm{Tr} }}

\def\nbar{\bar{n}}

\def\f12{\frac{1}{2}}

\newcommand{\mean}[1]{\ensuremath{\left\langle #1 \right\rangle}}

\newcommand {\comm}[2]{\left[ #1 , #2 \right]}

\begin{document}
\title{Universal decoherence due to gravitational time dilation}

\author{Igor Pikovski}\email{igor.pikovski@cfa.harvard.edu}
\affiliation{Vienna Center for Quantum Science and Technology (VCQ), University of Vienna, Boltzmanngasse 5, A-1090 Vienna, Austria}
\affiliation{Institute for Quantum Optics and Quantum Information (IQOQI), Austrian Academy of Sciences, Boltzmanngasse 3, Vienna A-1090, Austria}
\affiliation{ITAMP, Harvard-Smithsonian Center for Astrophysics, Cambridge, MA 02138, USA}
\affiliation{Department of Physics, Harvard University, Cambridge, MA 02138, USA}

 \author{Magdalena Zych}
\affiliation{Vienna Center for Quantum Science and Technology (VCQ), University of Vienna, Boltzmanngasse 5, A-1090 Vienna, Austria}
\affiliation{Institute for Quantum Optics and Quantum Information (IQOQI), Austrian Academy of Sciences, Boltzmanngasse 3, Vienna A-1090, Austria}
\affiliation{Centre for Engineered Quantum Systems, School of Mathematics and Physics, The University of Queensland, St Lucia, QLD 4072, Australia}

\author{Fabio Costa}
\affiliation{Vienna Center for Quantum Science and Technology (VCQ), University of Vienna, Boltzmanngasse 5, A-1090 Vienna, Austria}
\affiliation{Institute for Quantum Optics and Quantum Information (IQOQI), Austrian Academy of Sciences, Boltzmanngasse 3, Vienna A-1090, Austria}
\affiliation{Centre for Engineered Quantum Systems, School of Mathematics and Physics, The University of Queensland, St Lucia, QLD 4072, Australia}

\author{\v{C}aslav Brukner}
\affiliation{Vienna Center for Quantum Science and Technology (VCQ), University of Vienna, Boltzmanngasse 5, A-1090 Vienna, Austria}
\affiliation{Institute for Quantum Optics and Quantum Information (IQOQI), Austrian Academy of Sciences, Boltzmanngasse 3, Vienna A-1090, Austria}

\begin{abstract}
The physics of low-energy quantum systems is usually studied without explicit consideration of the background spacetime. Phenomena inherent to quantum theory on curved space-time, such as Hawking radiation, are typically assumed to be only relevant at extreme physical conditions: at high energies and in strong gravitational fields. Here we consider low-energy quantum mechanics in the presence of gravitational time dilation and show that the latter leads to decoherence of quantum superpositions. Time dilation induces a universal coupling between internal degrees of freedom and the centre-of-mass of a composite particle. The resulting correlations cause decoherence of the particle's position, even without any external environment. We also show that the weak time dilation on Earth is already sufficient to decohere micron scale objects. Gravity therefore can account for the emergence of classicality and the effect can in principle be tested in future matter wave experiments.
\end{abstract}
\maketitle

One of the most striking features of quantum theory is the quantum superposition principle. It has been demonstrated in numerous experiments with diverse systems, such as neutrons \cite{ref:COW1975}, atoms \cite{ref:Chu1991} and even large molecules \cite{ref:Eibenberger2013}. However, quantum superpositions are not observed on everyday, macroscopic scales. The origin of the quantum-to-classical transition is still an active field of research. A prominent role in this transition is commonly attributed to decoherence \cite{ref:GiuliniBook,ref:Zurek2003}:
due to interaction with an external environment, a particle gets entangled with its environment and loses its quantum coherence. Many specific models have been studied in which a particle interacts with its surrounding, such as a bath of phonons \cite{ref:Caldeira1983}, photons \cite{ref:Joos1985,ref:Hackermuller2004}, spins \cite{ref:Prokof'ev2000,ref:Hanson2008} and gravitational waves \cite{ref:Anastopoulos1996,ref:Lamine2006,ref:Blencowe2013}. An alternative route to explain classicality is taken in so-called wave function collapse models, which postulate an inherent breakdown of the superposition principle at some scale without any external environment \cite{ref:Penrose1996,ref:Diosi1989,ref:Bassi2013}. Such models are often inspired by general relativity but they rely on a fundamental modification of quantum theory. In contrast, here we derive the existence of decoherence due to time dilation without any modification of quantum mechanics and which takes place even for isolated composite systems. We show that even the weak time dilation on Earth is already sufficient to decohere micro-scale quantum systems.

We consider standard quantum mechanics in the presence of time dilation, with the focus on gravitational time dilation which causes clocks to run slower near a massive object. In Appendix A, we derive the Hamiltonian governing the quantum dynamics of a composite system on an arbitrary, static background space-time (and show that the same result is obtained as a limit of a quantum field model). Since we consider slowly moving particles and weak gravitational fields (i.e. to lowest order in $c^{-2}$), the results can also be obtained directly from the mass-energy equivalence \cite{ref:Einstein1911}: any internal energy contributes to the total weight of a system and thus also couples to gravity. Given any particle of mass $m$ and an arbitrary Hamiltonian $H_0$ that generates the time evolution of its internal degrees of freedom, gravity couples to the \textit{total} rest mass $m_{\textrm{tot}}=m+H_0/c^2$, i.e. gravity also couples to internal energy. The interaction with the gravitational potential $\Phi(x)$ is therefore $m_{\textrm{tot}} \Phi(x) = m\Phi(x) + H_{int} $, where $H_{int} = \Phi(x) H_0/c^2 $. This interaction term is just another formulation of gravitational time dilation (the same argument applies to inertial mass as well, so one recovers both the special relativistic and gravitational time dilation governed by $H_{int} = \Gamma(x,p) H_0/c^2$, with $\Gamma(x,p) = \Phi(x) -p^2/2m^2$).
For example, if the particle is a simple harmonic oscillator with frequency $\omega$, the above interaction with gravity effectively changes the frequency according to $\omega \rightarrow \omega(1+ \Phi(x)/c^2)$. This is the well-tested \cite{ref:Hafele1972,ref:Chou2010} gravitational redshift to lowest order in $c^{-2}$. When the energy is treated as a classical variable the time-dilation-induced interaction $H_{int}$ yields only this frequency shift. However, in quantum mechanics the internal energy $H_0$ and the position $x$ are quantized operators, thus time dilation causes an additional, purely quantum mechanical effect: entanglement between the internal degrees of freedom and the centre-of-mass position of the particle \cite{ref:Zych2011}.
Even though the time dilation on Earth is very weak, it leads to a significant effect for composite quantum systems, as we will show below.

Before deriving in full generality the time evolution of the centre-of-mass of an arbitrary composite system subject to time dilation, we consider a simplified model where a particle has $N/3$ constituents that are independent three-dimensional harmonic oscillators. Such a model equivalently describes $N$ internal harmonic modes of the particle. The internal Hamiltonian for this system is $H_0 = \sum_{i=1}^N \hbar \omega_i n_i $, where $n_i$ are the number operators for the $i$-th mode with frequency $\omega_i$. The centre-of-mass (with $x$ and $p$ being its vertical position and momentum, respectively) of the whole system is subject to the gravitational potential $\Phi(x)$.  For a homogeneous gravitational field in the x-direction we can approximate $\Phi(x) \approx g x$, where $g=9.81$~m/s$^2$ is the gravitational acceleration on Earth. The total Hamiltonian of the system is therefore $H~=~H_{cm}~+~H_0~+~H_{int}$, where $H_{cm}$ is some Hamiltonian for the centre-of-mass of the particle and the gravitational time-dilation-induced interaction (to lowest order in $c^{-2}$) between position and internal energy is
\be
H_{int} = \Phi(x)\frac{H_0}{c^2}   = \hbar \frac{g x}{c^2} \left(\sum_{i=1}^N  \omega_i n_i \right) \, .
\ee{eq:int}
To demonstrate decoherence, we first consider the case when the gravitational contribution to time dilation is dominant such that the velocity contributions can be neglected. A typical such case is a particle at rest in superposition of two vertically distinct positions $x_1$ and $x_2$ and a height difference $\Delta x =x_2 - x_1$. The centre-of-mass is in the state $\ket{\psi_{cm}(0)} = \frac{1}{\sqrt{2}}(\ket{x_1}+\ket{x_2})$. The internal degrees of freedom are in thermal equilibrium at local temperature $T$, thus each $i$-th constituent is described by the thermal density matrix $\rho_i = \left( \pi \nbar_i \right)^{-1} \int d^2 \alpha_i \exp(- |\alpha_i|^2/\nbar_i) \ket{\alpha_i}\bra{\alpha_i} $, where we used the coherent state representation with the average excitation $\nbar_i = (e^{\hbar \omega_i /k_B T}-1)^{-1}$ and where $k_B$ is the Boltzmann constant.  The total initial state is thus given by $\rho(0)=\ket{\psi_{cm}(0)} \bra{\psi_{cm}(0)} \otimes \prod^{N}_{i=1}  \rho_i$.
Gravitational time dilation now couples the centre-of-mass position of the system to the internal degrees of freedom $\rho_i$ via the Hamiltonian in eq.~\eqref{eq:int}. The off-diagonal elements $\rho_{12}=\bra{x_1}\rho \ket{x_2}=\rho_{21}^*$, which are responsible for quantum interference, evolve to $\rho_{12}(t)= \left( 2 \pi \nbar_i \right)^{-1} e^{i  m g \Delta x t/ \hbar} \times \prod^{N}_{i=1}  \int \! \! d^2 \alpha_i e^{- |\alpha_i|^2/\nbar_i}  \ket{\alpha_i e^{- i \omega_i(x_1) t}}\bra{\alpha_i e^{- i \omega_i(x_2) t}} $,
where $\omega_i(x) =\omega_i (1+ g x/c^2)$.
The frequencies of the internal oscillators depend on the position in the gravitational field, in accordance with gravitational time dilation (see also fig.~1).
To see decoherence of the centre-of-mass, we trace out the internal degrees of freedom. The quantum coherence can be quantified by the interferometric visibility $V(t) = 2 |\rho_{cm}^{(12)}(t)|= 2 |\Pi_{i=1}^N\Tr_i[ \rho_{12}(t) ] |$, which becomes $V(t) = \left| \prod^{N}_{i=1} \left[1+ \nbar_i\left(1- e^{-i \omega_i t g \Delta x/c^2}\right)\right]^{-1} \right|$.
This expression can be simplified for the typical case $\omega_i t g \Delta x/c^2 \ll 1$. In the high temperature limit we also have $\nbar_i \approx \frac{k_B T}{\hbar \omega_i}$, so that the frequency-dependence completely drops out from the visibility. In this case, the reduction of quantum interference is given by $V(t) \approx  \left( 1+ \left(\frac{k_B T g \Delta x t}{\hbar c^2}\right)^2 \right)^{-N/2}$. For times $t^2 \ll N \tau_{\textrm{dec}}^2$ this can be written as
\be
V(t) \approx  e^{-\left(t/\tau_{dec} \right)^2} \, ,
\ee{eq:D2}
where we defined the decoherence time
\be
\tau_{dec} = \sqrt{\frac{2}{N}} \frac{ \hbar c^2}{k_B T g \Delta x } \, .
\ee{eq:time}
The above equation shows that gravitational time dilation causes superpositions of composite systems to decohere.
The decoherence rate derived here scales linearly with the superposition size $\Delta x$, in contrast to other decoherence mechanisms that typically show a quadratic scaling \cite{ref:BreuerBook}. Also, decoherence due to gravitational time dilation depends on the number of oscillating internal states of the system, $N$. The suppression of quantum effects takes place even for completely isolated systems, provided that the superposition amplitudes acquire a sufficient proper time difference. In the high temperature limit the frequencies of the internal oscillations drop out entirely from the final expression, therefore it is not necessary to have fast-evolving internal states.  Note that the decoherence derived here depends on the constants $\hbar$, $c$, $k_B$ and the gravitational acceleration $g$: it can therefore be considered a relativistic, thermodynamic and quantum mechanical effect.

\begin{figure}
\includegraphics[width=0.99\columnwidth]{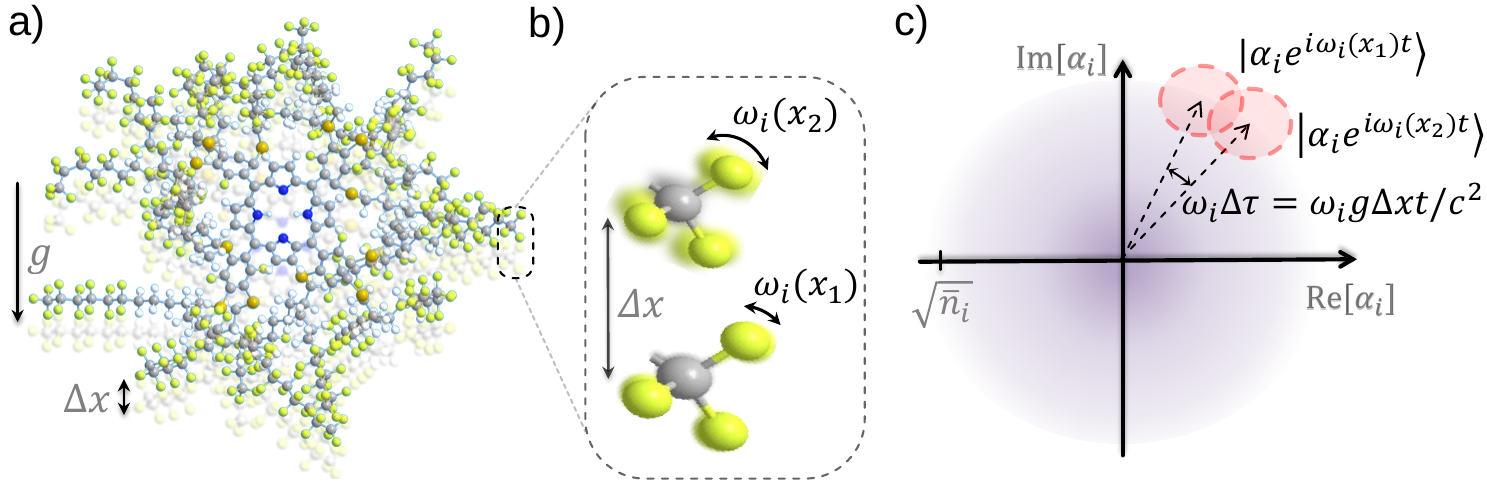}
\caption{\small Gravitational time dilation causes decoherence of composite quantum systems. a) Illustration of a TPPF20 molecule which has recently been used for matter-wave interference \cite{ref:Eibenberger2013}. Here we illustrate a vertical superposition of size $\Delta x$ in Earth's gravitational potential $\Phi(x)=gx$. b) The frequencies $\omega_i$ of internal oscillations are modified in the gravitational field, i.e. $\omega_i \rightarrow \omega_i(x)=\omega_i(1+gx/c^2)$, which correlates the internal states and the centre-of-mass position of the molecule. c) Phase space representation of the $i$-th constituent which is in a thermal state with average occupation $\bar{n}_i \approx k_B T/\hbar \omega_i$. In the coherent state representation of the internal states, the frequency of each coherent state depends on the position of the whole molecule and thus differs between the two superposed amplitudes by an amount $\omega_i \Delta \tau$. Even for small time-dilations, this causes decoherence of the molecule with N constituents after a time $\tau_{dec}$, given in eq.~\eqref{eq:time}. }
\end{figure}

The effect is very general and originates from the total proper time difference between superposed world lines. Quantum systems with internal degrees-of-freedom are affected on arbitrary space-time metrics, as long as a proper time difference is accumulated. To highlight this, consider a particle moving in superposition along two arbitrary world lines with proper time difference $\Delta \tau$ (see fig.~2). The two superposed amplitudes can interfere when the world lines meet. Due to time dilation, the internal energy is effectively altered by $H_0(1+ \Gamma/c^2)$, with $\Gamma(x,p)= \Phi(x)-p^2/2m^2$ (see also Appendix A). Each amplitude therefore evolves with $U(t)=\exp[-\frac{i}{\hbar} \int dt(H_{cm}+(1+\frac{\Gamma}{c^2})H_0)]$ along the respective world lines. $H_0$ does not depend on $x$, $p$ and $t$ and for clarity, we restrict the analysis to semi-classical paths, i.e. constrained to have coordinates $\bar{x}_{1}(t)$, $\bar{p}_{1}(t)$ and $\bar{x}_{2}(t)$, $\bar{p}_{2}(t)$ along the two world lines, respectively (as in fig.~2c). The interference visibility is then $V =  \left| \Tr[e^{-\frac{i}{\hbar} \int dt(1+\Gamma(\bar{x}_{1}, \bar{p}_{1})/c^2) H_0}\rho_0 e^{\frac{i}{\hbar} \int dt(1+\Gamma(\bar{x}_{2}, \bar{p}_{2})/c^2) H_0}] \right|$. Since $d\tau = dt \sqrt{g_{\mu \nu} \dot{x}^{\mu}\dot{x}^{\nu}} \approx dt(1+\Gamma/c^2)$, the interference visibility is simply
\be
V =  \left| \mean{e^{-i H_0 \Delta \tau/\hbar}} \right| \, ,
\ee{eq:visibility}
where $\Delta \tau = \tau_1 - \tau_2$ is the proper time difference between the two world lines and the expectation value is taken with respect to the initial state. This result is manifestly coordinate invariant and shows that decoherence occurs if a proper time difference is present, and if the internal states are not eigenstates of internal energy. Eq.~\eqref{eq:time} is recovered as a special case of the general formula above (expanding to lowest order in $\Delta \tau$, assuming N internal harmonic oscillators such that $\Delta E_0^2 = \mean{H_0^2} - \mean{H_0}^2 \approx N k_B^2 T^2$ and neglecting the $p$-dependent term yields eq.~\eqref{eq:time}). For the special case of a pure two-level system, eq.~\eqref{eq:visibility} also reproduces the effect discussed in ref.~\cite{ref:Zych2011}, which can be interpreted as due to the which-way information acquired by a clock. In contrast, which-way information is never available for thermal states (a more detailed discussion can be found in Appendix C). Eq.~\eqref{eq:visibility} shows, however, that time dilation affects any state that is not an eigenstate of $H_0$. The coupling is universal, which follows directly from the universality of time dilation. Thus the decoherence is as universal as time dilation itself, in the sense that all composite quantum systems are affected, independently of the nature and kind of their internal energy $H_0$.
\begin{figure}[!ht]
\includegraphics[width=0.99\columnwidth]{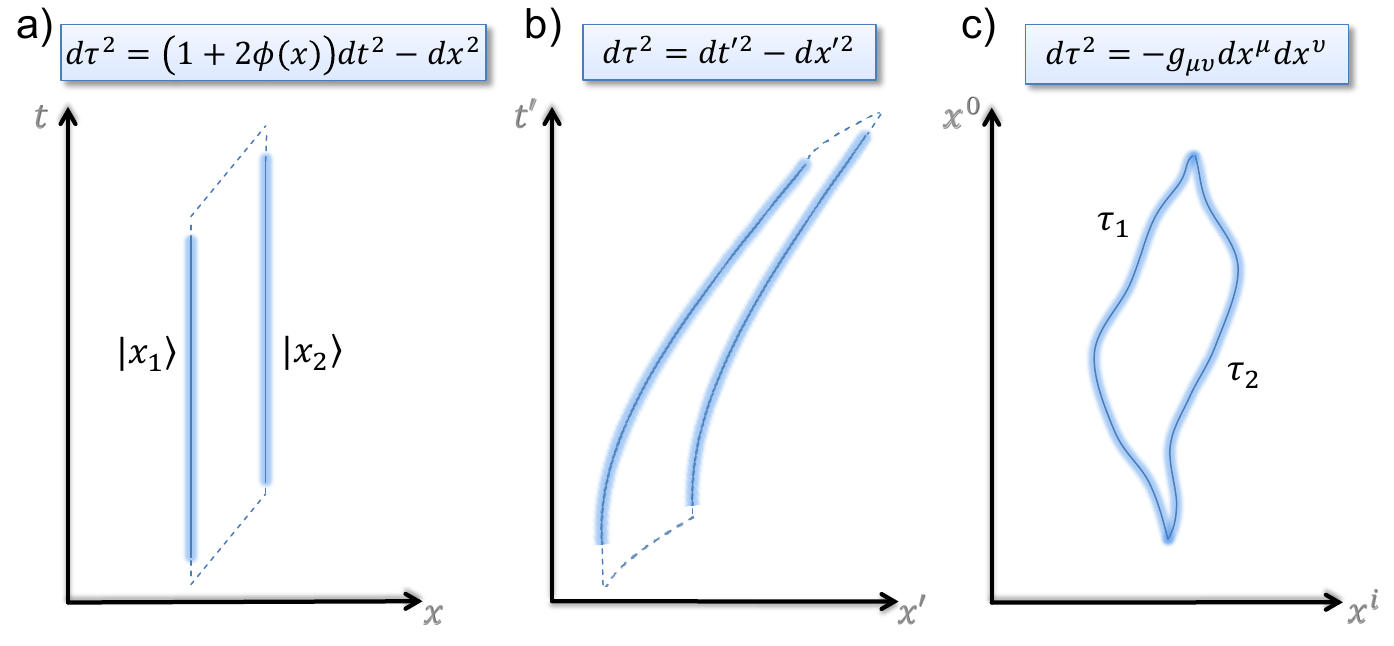}
\caption{\small A composite particle in superposition will decohere due to time dilation. The figure shows two superposed world lines in different situations and space-times (here $c=1$). a) A particle in superposition at two different fixed heights above the Earth, as considered in the main text (the dashed lines represent arbitrary small non-stationary contributions necessary to perform an interferometric experiment). The centre-of-mass will decohere after a time $\tau_{dec} $ as given in eq.~\eqref{eq:time}. In general, the full evolution of the centre-of-mass is given by eq.~\eqref{eq:MEfull}.  b) A particle undergoing uniform acceleration $g$ in flat space-time will experience the same time dilation and thus the same decoherence as in case a). Equivalently, the diagram describes the previous situation from the point of view of a freely-falling observer. c) A composite particle with internal Hamiltonian $H_0$ in an arbitrary space-time will decohere if the two superposed trajectories differ in proper time. The visibility of quantum interference $V$ reduces depending upon on the proper time difference $\Delta \tau$, see eq.~\eqref{eq:visibility}.}
\end{figure}

We now consider the full time evolution in the presence of time dilation that takes any arbitrary internal Hamiltonian $H_0$ and centre-of-mass Hamiltonian $H_{cm}$ into account. To this end, we derive a master equation that describes the quantum dynamics of a composite system on a background space-time to lowest non-vanishing order in $c^{-2}$ (see Appendix B for details). The resulting time evolution of the centre-of-mass in the presence of special relativistic and gravitational time dilation is
\be
\begin{split}
 & \dot{\rho}_{cm}(t)  = \!  -\frac{i}{\hbar}\comm{H_{cm} +  \frac{\bar{E}_0 }{ c^2} \Gamma(x,p)}{\rho_{cm}(t)} -  \left( \frac{\Delta E_0 }{ \hbar c^2} \right)^{\! 2} \times  \\
&  \! \! \times \! \! \int_0^{t} \! \! ds \comm{\Gamma(x,p)}{e^{- i H_{cm} s/\hbar} \comm{\Gamma(x,p)}{\rho_{cm}(t-s)} e^{ i H_{cm}s /\hbar}} ,
\end{split}
\ee{eq:MEfull}
where $\Gamma(x,p)= \Phi(x)-p^2/2m^2$.
The first term describes the unitary evolution of the centre-of-mass due to an arbitrary Hamiltonian $H_{cm}$, which is completely general and can also include external interactions (as for example those necessary for keeping the particle in superposition or realizing an interference experiment) as well as relativistic corrections to the centre-of-mass dynamics. The correction dependent upon $\bar{E}_0 = \mean{H_0}$ stems from the relativistic contribution to the mass.
The second term causes the suppression of off-diagonal elements of the density matrix and is responsible for the decoherence. It is proportional to $\Delta E_0^2=\mean{H_0^2}-\mean{H_0}^2 $, the fluctuations in internal energy, or equivalently the heat capacity $C_v = \Delta E_0^2/k_B T^2$ (the high temperature limit of the model that we used previously corresponds to the Einstein solid model).
The integral captures the fact that decoherence depends on the overall acquired proper time difference during a particle's evolution.
For a stationary particles, and if the centre-of-mass Hamiltonian $H_{cm}$ does not induce significant changes to the off-diagonal elements on the decoherence time scale,
the master equation becomes approximately
\be
\begin{split}
 \dot{\rho}_{cm}(t)  & \approx -\frac{i}{\hbar}\comm{\tilde{H}_{cm} +  \left( m +  \frac{\bar{E}_0}{ c^2} \right)gx}{\rho_{cm}(t)} - \\
& -  \left( \frac{ \Delta E_0 g}{ \hbar c^2} \right)^2 t \comm{x}{\comm{x}{\rho_{cm}(t)}} .
\end{split}
\ee{eq:ME}
In the unitary part we have separated for clarity the Newtonian gravitational potential (i.e. $H_{cm} = \tilde{H}_{cm}+mgx$): It is evident that the potential couples to an effective total mass $m_{\textrm{tot}}=m+\bar{E}_0/c^2$ that includes the average internal energy, which becomes $\bar{E}_0= \mean{H_0} \approx N k_B T$ for the previously considered model with $N$ thermal internal harmonic oscillators. This is in accordance with the notion of heat in general relativity (in Einstein's words \cite{ref:EinsteinBook}: ``a piece of iron weighs more when red-hot than when cool''), the relation to the Tolman effect is briefly discussed in Appendix C.
The non-unitary part now depends only on the stationary $x$-contributions (see also fig.~2a).
The decoherence time scale is found from the solution to eq.~\eqref{eq:ME}, which for the off-diagonal terms $\rho_{cm}^{(12)}$ is approximately (to order $O(\hbar^{-2})$): $\rho_{cm}^{(12)}(t) \sim \rho_{cm}^{(12)}(0) \, e^{- (t/\tau_{dec})^2}$, with $\tau_{\textrm{dec}}= \sqrt{2} \hbar c^2/ (\Delta E_0 g  \Delta x)$.
The loss of visibility thus agrees with eq.~\eqref{eq:D2} and the decoherence time scale reduces to eq.~\eqref{eq:time} for the specific model used previously. The master equation due to gravitational time dilation, eq.~\eqref{eq:ME}, is similar in form to  other master equations typically studied in the field of decoherence \cite{ref:GiuliniBook,ref:BreuerBook} but does not include any dissipative term. Thus time dilation provides naturally an ``ideal'' master equation for decoherence that suppresses off-diagonal terms in the position basis for stationary particles. For non-stationary systems, decoherence is governed by eq.~\eqref{eq:MEfull} and the pointer basis derives from a combination of $x$ and $p$. Position and momentum are therefore naturally driven into becoming classical variables. The evolution in the presence of gravitational time dilation is inherently non-Markovian, since the overall acquired proper time difference is crucial. This results in a Gaussian decay (rather than an exponential decay as in Markovian models) of the off-diagonal elements and the decoherence time directly depends on the fluctuations in internal energy. This again highlights the interplay between thermodynamics, relativity and quantum theory that is relevant for this effect.

To estimate the strength of the decoherence due to time dilation, we make use of eq.~\eqref{eq:time} and consider a human-scale macroscopic system at room temperature. Assuming that the system has Avogadro's number of constituent particles which oscillate, we set $N \sim 10^{23}$, which amounts to a gram-scale system. For a superposition size of $\Delta x=10^{-6}$~m, the decoherence time \eqref{eq:time} becomes
\be
\tau_{\textrm{dec}}  \approx 10^{-3} s  \, .
\ee{eq:decTime}
Remarkably, even though the gravitational time dilation is very weak, its resulting decoherence is already substantial on human scales and not just for astrophysical objects. Macroscopic objects completely decohere on Earth on a short time scale due to gravitational time dilation. In contrast to other decoherence mechanisms, this effect cannot be shielded and decoherence will occur whenever there is time dilation between superposed amplitudes. But as any other decoherence in quantum theory, the effect is  in principle reversible: as is apparent from eq.~ \eqref{eq:visibility}, ``revivals'' of coherence will occur after a sufficiently large proper time difference is accumulated, dependent on the frequencies of the internal degrees of freedom. However, the corresponding recurrence times typically scale exponentially with the size of the system \cite{ref:Hogg1982}, thus time dilation can cause decoherence which is irreversible ``for all practical purposes'' if interfering paths have a proper time difference. This will be increasingly difficult to control in experiments with large, composite systems, and is simply unavoidable for systems not under experimental control.

From the perspective of quantum theory, decoherence due to time dilation is fully analogous to any other decoherence source. The loss of coherence takes place due to correlations with degrees of freedom that are not accessible, here internal degrees of freedom of the composite system. The unique aspect of the effect described here is that the correlations are induced by relativistic time dilation, and would not take place in Newtonian gravity. Thus, to understand this effect it is necessary to invoke quantum theory and time dilation stemming from Earth's gravitational field. The phenomenon arises already in weak, stationary space-times and decoheres composite systems into the position basis, even if they are isolated from any external environment. It is thus of different nature than decoherence due to scattering with gravitational waves \cite{ref:Anastopoulos1996,ref:Lamine2006,ref:Blencowe2013}.
Importantly, the time-dilation-induced decoherence is entirely within the framework of quantum mechanics and classical general relativity. No free ``model parameters'' enter and unitarity is preserved on a fundamental level. This is in stark contrast to hypothetical models where gravity leads to spontaneous collapse of the wave function and that require a breakdown of unitarity \cite{ref:Penrose1996,ref:Diosi1989,ref:Bassi2013} or include stochastic fluctuations of the metric \cite{ref:Karolyhazy1966}. Our results show that general relativity can account for the suppression of quantum behavior for macroscopic objects without introducing any modifications to quantum mechanics or to general relativity.

In astrophysical settings, the decoherence can even be substantially stronger: the time scale \eqref{eq:time} can be rewritten in terms of the Schwarzschild-radius $R_s=2 G M /c^2$ of the background space-time as $\tau_{\textrm{dec}} =  \sqrt{8/N} (\hbar R^2/k_B TR_s \Delta x)$, where $R$ is the distance between the particle and the centre of the gravitating object (since the nature of the effect is gravitational, quantum mechanical and thermodynamic, the decoherence time may also be written in terms of the Hawking-temperature $T_H=\hbar c^3 / 8 \pi k_B G M$ of the body with mass $M$, but the decoherence is not related to any horizon and the appearance of the Hawking temperature is solely a reformulation of the fundamental constants involved). The decoherence is stronger for high masses $M$ and for small distances $R$ to the mass, i.e. for stronger time dilation. At the horizon of a black hole with 5 solar masses, a nm-size superposition of a gram-scale object at $T=1~$K would decohere after about $\tau_{\textrm{dec}} \approx 1~$ns.

\begin{figure}[t]
\includegraphics[width=0.99\columnwidth]{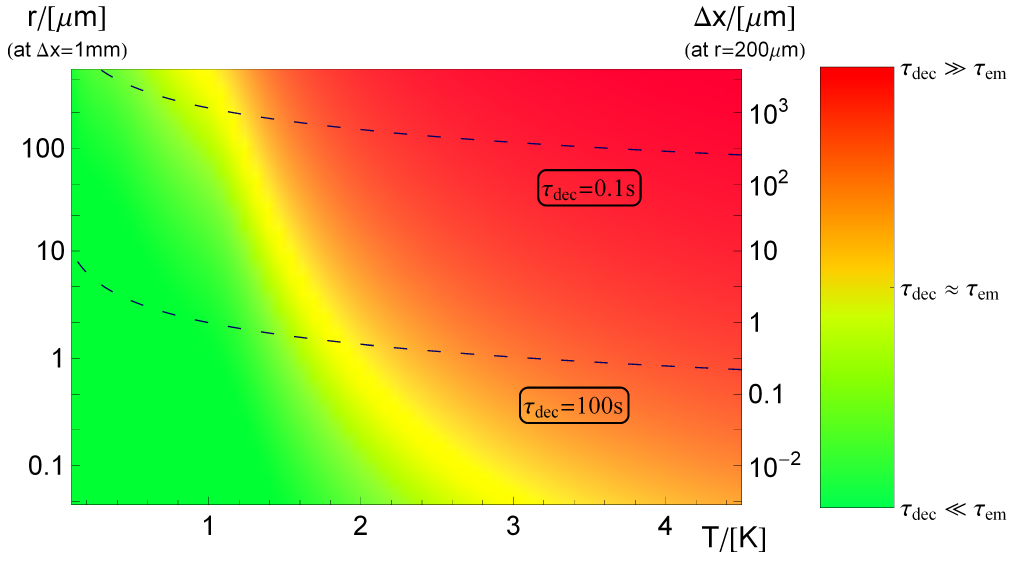}
\caption{\small Decoherence due to gravitational time dilation as compared to decoherence due to emission of thermal radiation for sapphire microspheres. In the green region time dilation is the dominant decoherence mechanism. The left axis shows various sphere radii $r$ (corresponding to particle numbers $N=10^7$ to $N=10^{18}$) for a fixed superposition size $\Delta x$, whereas the right axis shows various superposition sizes for a fixed particle radius. The dashed lines correspond to the respective time dilation decoherence time scales as in eq.~\eqref{eq:time}. Sapphire was chosen for its low emission at microwave frequencies.
} \label{fig:DecParams}
\end{figure}

We now discuss a possible direct experimental verification of the derived decoherence mechanism. The gravitational time dilation is well tested in classical physics \cite{ref:Hafele1972,ref:Chou2010} but the quantized Hamiltonian \eqref{eq:int} has not yet been studied experimentally. In particular, an experiment to study the induced quantum entanglement of internal degrees of freedom with the centre-of-mass mode, first proposed in ref.~\cite{ref:Zych2011}, has not yet been realised. To confirm this quantum mechanical interaction one can use controllable internal states in matter wave interferometry \cite{ref:Zych2011}, or use Shapiro-delay in single photon interference \cite{ref:Zych2012} (a related effect for entangled photon pairs was also discussed in ref.~\cite{ref:Ralph2006}). Such an experimental verification of the quantum Hamiltonian \eqref{eq:int} would be a strong indication for the presence of the decoherence described here.
To test directly the decoherence due to time dilation it is necessary to bring relatively complex systems into superposition. This can in principle be achieved with molecule interferometry \cite{ref:Eibenberger2013,ref:Hackermuller2004}, cooled microspheres \cite{ref:Kiesel2013,ref:Asenbaum2013} or with micro-mechanical mirrors \cite{ref:Marshall2003}. The latter, however, is expected to be limited to very small separations only (on the order of 1~pm) and is therefore less suitable. To see decoherence caused by time dilation, other decoherence mechanisms will need to be suppressed: The scattering with surrounding molecules and with thermal radiation requires such an experiment to be performed at liquid Helium temperatures and in ultra-high vacuum \cite{ref:Joos1985}. Additionally, the emission and absorption of thermal radiation by the system \cite{ref:Hackermuller2004} will be a competing decoherence source. For the parameter regime studied here, emission of radiation is expected to be the dominant decoherence effect, with the decoherence time \cite{ref:GiuliniBook,ref:Joos1985} $\tau_{\textrm{em}}= (\int dk k^2 c \, g(k)  \sigma_{\textrm{eff}}(k) \Delta x^2)^{-1} $, where $g(k)$ is the mode density of the wave vectors $k$ and $\sigma_{\textrm{eff}}(k)$ the effective scattering cross section. To see the time-dilation-induced decoherence, we require that the decoherence due to emission of radiation is weaker than due to time dilation, i.e. $\tau_{\textrm{dec}} \lesssim \tau_{\textrm{em}}$. To get quantitative estimates, we rely on the harmonic oscillator model introduced previously and show in fig.~3 the parameter regime where time-dilation-induced decoherence can in principle be distinguished from decoherence due to thermal emission, focussing on micro-scale particles at cryogenic temperatures (sapphire was chosen due to its low microwave emission at low temperatures \cite{ref:Lamb1996}). The emission of radiation can be further suppressed if the mode density is reduced, which can ease the restrictions on temperature. However, we note that the simple model for the composition of the system, necessary to estimate the time dilation decoherence rate, is very crude and at low temperatures we expect the model to break down. Given a specific system, the time dilation decoherence can be estimated more accurately by measurement of the internal energy fluctuations through the heat capacity. Although an experiment to measure decoherence due to proper time is very challenging, the rapid developments in controlling large quantum systems \cite{ref:Eibenberger2013,ref:Kiesel2013,ref:Asenbaum2013} for quantum metrology and for testing wave function collapse models \cite{ref:Bassi2013,ref:Marshall2003} will inevitably come to the regime where the time-dilation-induced decoherence predicted here will be of importance. In the long run, experiments on Earth will have to be specifically designed to avoid this gravitational effect on quantum coherence. As a final remark, we note that due to the universality of time dilation, all dynamical processes contribute to this decoherence, even those that are typically experimentally inaccessible (such as nuclear dynamics, which has not been taken into account in our treatment). Thus time dilation decoherence could also serve as a tool to indirectly study dynamical processes within composite systems.

\begin{acknowledgments}
We thank M. Arndt, M. Aspelmeyer, L. Diosi and M. Vanner for discussions and S. Eibenberger for providing us with the illustration of the TPPF20 molecule. This work was supported by the the Austrian Science Fund (FWF) through the doctoral program Complex Quantum Systems (CoQuS), the Vienna Center for Quantum Science and Technology (VCQ), the SFB FoQuS and the Individual Project 24621, by the Foundational Questions Institute (FQXi), the John Templeton Foundation, the Australian Research Council Centre of Excellence for Engineered Quantum Systems (grant number CE110001013), the European Commission through RAQUEL (No. 323970) and the COST Action MP1209.
\end{acknowledgments}

\pagebreak

\section*{Appendix}
\subsection{Hamiltonian for gravitational time dilation}
We present a Hamiltonian formalism useful to describe the dynamics of low energy quantum systems with internal degrees of freedom subject to time dilation. We further show an explicit example of a quantum field in curved space-time whose dynamics reduces to the same Hamiltonian in the appropriate limit.

We consider the time evolution of a composite quantum system in the low-energy limit on a generic space-time, described by a metric $g_{\mu \nu}$ with signature ($-+++$). We restrict the treatment to static metrics with $\partial_0 g_{\mu \nu}=0$ and $g_{0i}=0$, where Latin indices refer to the spatial 3-components. The typical systems we consider are low energy quantum systems (such as atoms, molecules, nanospheres, etc.) with internal energy levels (such as electronic, rotational, vibrational) in a weak gravitational field and under small accelerations. For such systems, one can assume that the relative distances between their constituents are sufficiently small, such that variations of the metric over their extension can be neglected. In this case one can assign a single position degree of freedom to the centre-of-mass of the system, which in the classical limit describes a single world line. In other words, we consider the limit in which the system can be effectively considered as ``point-like'' with internal degrees of freedom. This is directly analogous to the notion of ideal clocks in relativity, which measure time along a well-localized world line.

The rest energy $E_{\text{rest}}$ of the system is defined as the invariant quantity
\be
p_{\mu} p^{\mu} = g^{\mu \nu} p_{\mu} p_{\nu}=-(E_{\text{rest}}/c)^2 \, ;
\ee{eq:p2}
it corresponds to the total mass-energy as measured by a co-moving observer. Here $p_{\mu}$ is the system's total 4-momentum in arbitrary coordinates (we restrict to coordinates which keep the stated assumptions for the metric).
The dynamics of the system can be described in terms of the evolution with respect to an arbitrary time coordinate $t$. The generator of the coordinate-time translations follows from eq.~\eqref{eq:p2} and is given by the Hamiltonian
\begin{equation}
H= cp_0 = \sqrt{-g_{00}\left(E_{\text{rest}}^2 + c^2 g^{ij} p_i p_j \right)}.
\label{Hamiltonian}
\end{equation}
If the particle is at rest with respect to a static observer, the energy is $H=\sqrt{-g_{00}}E_{\text{rest}}$. This ``redshift factor'' is also sometimes expressed in terms of the time-like killing vector $k^{\mu}$ as $\sqrt{-k_{\mu} k^{\mu}}=\sqrt{-g_{00}}$. The rest energy $E_{\text{rest}}$ is the total Hamiltonian of the system in its local comoving frame. The static rest mass contribution $mc^2$ can be explicitly separated, and the remaining part is just the Hamiltonian of the internal degrees of freedom, which we denote by $H_0$. The full dynamics of internal and external degrees of freedom is thus governed by the total Hamiltonian \eqref{Hamiltonian}, with
\begin{equation}
E_{\text{rest}}=mc^2+H_0 \, .
\label{split}
\end{equation}
Relativistically, there is no distinction between ``rest mass'' and ``rest energy''. In fact, the largest contribution to the rest mass $m$ of the systems we consider, e.g.\ a molecule, is already given by binding energies between atoms, nucleons, quarks and all other constituents (at an even more fundamental level, masses of fundamental particles are reducible to interaction energies with the Higgs field, according to the standard model of particle physics). The natural choice for the split \eqref{split} is dictated by the energy scale: if some degrees of freedom are ``frozen'', their contribution to the rest energy can be incorporated in the mass term. The split between mass and internal energy \eqref{split} is thus merely conventional, and amounts to a choice of the zero-energy state of the internal degrees of freedom.
The general expression \eqref{split} can also be derived from the action of a particle in the comoving frame, in which the time coordinate coincides with the proper time: $S=\int L_{\text{rest}} d\tau$, where $L_{\text{rest}}=L_{\text{rest}}(q_i, q_i')$ is the Lagrangian describing the internal degrees of freedom with coordinates $q_i$ and $q_i'=dq_i/d\tau$. Changing to the lab frame this expression becomes $S=\int L_{\text{rest}}(q_i, \dot{q}_i t' ) \dot{\tau} dt$, with $\dot{\tau}=d\tau/dt=\sqrt{g_{\mu \nu }\dot{x}^{\mu}\dot{x}^{\nu}}$ and $t'=dt/d\tau$. The Legendre transform yields the Hamiltonian for the system, which gives exactly the expression \eqref{Hamiltonian} with
\be
E_{\text{rest}} = \frac{\partial L_{\text{rest}}}{\partial q_i'} q_i' - L_{\text{rest}} .
\ee{eq:Erest}
If the internal dynamics is irrelevant, we simply have $L_{\text{rest}}=-mc^2$ and $E_{\text{rest}}=mc^2$. But in general any arbitrary internal dynamics governed by $L_{\text{rest}}$ gives rise to the total energy as in eq.~\eqref{Hamiltonian} with $E_{\text{rest}}$ as in eq.~\eqref{eq:Erest}. For example, two masses on a spring (with total mass $m$, reduced mass $\mu$ and spring constant $k$) in the comoving frame of the centre-of-mass are described by $L_{\text{rest}}=\mu q'^2/2 - kq^2/2 - mc^2$, where $q$ is the relative degree of freedom of the two masses. This gives $E_{\text{rest}}=mc^2 + \mu q'^2/2 + kq^2/2 = mc^2 + H_0$, where $H_0$ describes the dynamical part of the internal degrees of freedom.

To obtain the quantum equations of motion one can replace the 4-momenta in the full expression \eqref{eq:p2} with covariant derivatives, which leads to a modified Klein-Gordon equation with $E_{\text{rest}}$ as the invariant total mass (a similar derivation can be applied to the Dirac equation to describe particles with spin). For energies small compared to $E_{\text{rest}}$ (such that particle creation and other quantum field effects are negligible), the Klein-Gordon field is treated as a particle in first quantization and the low-energy Schr\"odinger evolution is obtained. Specifically, the Hamiltonian for the dynamics of a point particle on a post-Newtonian background metric  with $g_{00} = - (1+ 2 \Phi(x)/c^2+2\Phi^2(x)/c^4)$ and $g_{ij}=\delta_{ij}(1-2 \Phi(x)/c^2)$ is obtained following ref.~\cite{ref:Laemmerzahl1996} (with the addition of internal degrees of freedom in $E_{\text{rest}}$), resulting in
\be
\begin{split}
H  =& \bar{m}_rc^2 + \frac{p^2}{2\bar{m}_r}  + \bar{m}_r\Phi(x) - \frac{p^4}{8\bar{m}_r^3c^2} +   \frac{\bar{m}_r\Phi^2(x)}{2c^2}  \\
&+ \frac{3}{2\bar{m}_rc^2}\left(\Phi(x) p^2 + \left[ p \Phi(x) \right] p + \f12 \left[ p^2\Phi(x)\right] \right) ,
\end{split}
\ee{eq:HLamm}
where $\left[ p \Phi \right]$ acts only on the potential and we introduced $\bar{m}_r:= E_{\text{rest}}/c^2$ to keep the expansion to order $c^{-2}$ explicit (in deriving the Hamiltonian there is an ambiguity in the ordering of the $p \Phi(x)$ terms, but which does not affect the results for time dilation).
For an eigenstate $\ket{E_i}$ of the internal Hamiltonian $H_0$ the rest energy in the Hamiltonian \eqref{eq:HLamm} can be treated as a parameter $\bar{m}_r =m+E_i/c^2$. For arbitrary internal states and due to the linearity of quantum mechanics, $E_{\text{rest}}$ has to be treated as an operator acting on the internal degrees of freedom according to \eqref{split}. The Hamiltonian \eqref{eq:HLamm} thus describes the full quantum dynamics of the system, including internal and external degrees of freedom. Expanding the result to first order in $H_0/mc^2$, we find
\be
\begin{split}
H   & = H_{cm} + H_0 \left( 1+ \frac{\Phi(x)}{c^2}- \frac{p^2}{2 m c^2} \right)  \\
& = H_{cm} + H_0 \left( 1+ \frac{\Gamma(x,p)}{c^2} \right),
\end{split}
\ee{eq:Hfull}
where  $H_{cm}$ includes all terms acting on the centre-of-mass to this order of approximation (and can also include any other interaction, such as the electromagnetic interaction \cite{ref:Laemmerzahl1996,ref:Kiefer1991}). The term $\Gamma(x,p)=\Phi(x) - p^2/2m^2$ captures the time dilation, which effectively shifts the internal energy. An alternative route to get eq.~\eqref{eq:Hfull} is to directly expand eq.~\eqref{Hamiltonian} in orders of $c^{-2}$ and canonically quantize the result, which yields the same Schr\"odinger equation with relativistic corrections. The term proportional to $\Phi(x)$ stems from $\sqrt{-k_{\mu} k^{\mu}}=\sqrt{-g_{00}}$, while the $p^2$-term stems from the spatial $g_{ij}$ components of the metric. The former captures gravitational time dilation and the latter is the velocity-dependent special relativistic time dilation. In total, eq.~\eqref{eq:Hfull} describes the special and general relativistic corrections to the dynamics of a quantum system with internal degrees of freedom, to lowest order in $c^{-2}$. Note that the coupling between internal and external degrees of freedom is completely independent of the nature and kinds of interactions involved in the internal dynamics $H_0$. This is a consequence of the universality of time dilation, which affects all kinds of clocks, irrespectively of their specific construction.

The description above is well-suited for low-energy particles on a background space-time. The same results can also be obtained from a field theory description and to highlight this, we consider an explicit example in which the effective Hamiltonian is derived starting from a quantum field. We consider a Klein-Gordon field of mass $m$, in a state approximately localized in some region of space. The position of the region of space represents the ``centre-of-mass'', while the field describes the ``internal degrees of freedom'' of our system. The dynamics of the field is given by the action
\begin{equation}
\label{curvedKG}
S=-\frac{1}{2}\int{d^4x \sqrt{-g}\left[g^{\mu \nu}\frac{\partial \phi}{\partial x^{\mu}}\frac{\partial \phi}{\partial x^{\nu}} + m^2 \phi^2\right]},
\end{equation}
where $g$ is the determinant of $g_{\mu \nu}$ and we use units with $c=1$. To simplify the discussion, we consider a system at a fixed height above Earth. The effect already appears by expanding the metric at the first order: $g_{ij}=\delta_{ij}$, $g_{00} = -\left[1+2\Phi(x)\right]$, where $\Phi(x)$ is the Newtonian potential. In this case the determinant is $g= g_{00}$ and $g^{00}=1/g_{00}$, so we can rewrite eq.~\eqref{curvedKG} as
\begin{equation}
S=-\frac{1}{2}\int{d^4x \sqrt{-g_{00}}\left[\frac{\left(\partial_t \phi\right)^2}{g_{00}}+ \left| \bf{\nabla} \phi \right|^2 + m^2 \phi^2\right]}.
\end{equation}
Once the coordinates are fixed, we can write $S=\int dt L$ and single out the Lagrangian
\begin{equation}
\label{lagrangian}
L=\frac{1}{2}\int{d^3x \left[\frac{\left(\partial_t \phi\right)^2}{\sqrt{-g_{00}}} - \sqrt{-g_{00}} \left(\left| \bf{\nabla} \phi \right|^2 + m^2 \phi^2\right)\right]}.
\end{equation}
In order to pass to the Hamiltonian picture, we need the conjugate momenta
\begin{equation}
\pi(x)= \frac{\delta L}{\delta \partial_t \phi(x)} = \frac{1}{\sqrt{-g_{00}}} \partial_t \phi(x).
\label{momenta}
\end{equation}
The Hamiltonian in these coordinates is given by $H= \int{d^3x\, \pi(x) \partial_t \phi(x)} - L$. Substituting $\partial_t \phi(x) = \sqrt{-g_{00}}\pi(x)$, we get
\begin{equation}
\begin{split}
H =& \frac{1}{2}\int{d^3 x} \sqrt{-g_{00}} \left[ \pi(x)^2 + \left(\left| \bf{\nabla} \phi \right|^2 + m^2 \phi^2\right)\right].
\end{split}
\label{fieldHam}
\end{equation}
We are in particular interested in a system of small size sitting at a fixed space coordinate $\bar{x}$ above Earth. This can be modelled by confining the field to a small box of volume $V$ around $\bar{x}$, so that the space integral \eqref{fieldHam} can be restricted to that volume. If the potential is approximately constant within the volume, $\Phi(x) \approx \Phi(\bar{x})$ for $ x\in V$, we can take it out of the integral and obtain
\begin{equation}
H \approx \sqrt{-g_{00}(\bar{x})} H_{0} = \sqrt{1+2\Phi(\bar{x})} H_0,
\label{smallV}
\end{equation}
where the \textit{rest Hamiltonian} is given by $H_{0}= \frac{1}{2}\int_V d^3 x\left[ \pi(x)^2 + \left| \bf{\nabla} \phi \right|^2 + m^2 \phi^2\right]$, which is just the usual Klein-Gordon Hamiltonian in a finite volume in Minkowski space-time. The factor in front of $H_0$ in eq.\ \eqref{smallV} is responsible for the gravitational red-shift: all energies, when measured according to coordinate time, are rescaled with respect to those measured locally at the position of the system. We can see this explicitly by considering a cubic box of side $l=V^{\frac{1}{3}}$, for which the rest Hamiltonian can be diagonalized as
\begin{equation}
H_{0} = \sum_{\bf{k}} E_\textbf{k}a_{\textbf{k}}^{\dag}a_{\textbf{k}},
\end{equation}
where, restoring units, $E_\textbf{k} = c\sqrt{\textbf{k}^2 + m^2c^2}$, $k_j=\frac{2 \pi \hbar}{l} n_j$ for $n_j \in \mathbb{N}$, $j=1,2,3$ and $a_{\textbf{k}}^{\dag}$ ($a_{\textbf{k}}$) creates (annihilates) a boson with momentum $\textbf{k}$ (neglecting the constant vacuum energy). For a particle at a distance $x$ above Earth, we can set
$\sqrt{-g_{00}(x)} \approx 1+\Phi(x)/c^2 \approx 1+gx/c^2$, where $g$ is the gravitational acceleration. The Hamiltonian \eqref{smallV} thus becomes
\begin{equation}
H \approx \left(1+\frac{gx}{c^2}\right) \sum_{\bf{k}} E_\textbf{k} a_{\textbf{k}}^{\dag}a_{\textbf{k}}.
\label{coupling}
\end{equation}
Equations \eqref{coupling} shows the same coupling between position and internal energy as described in the main text (eq.~\eqref{eq:int}). It is valid for a system at rest in the chosen coordinate system, i.e.\ with vanishing external momentum.
The contribution of the external momentum is recovered by moving to an arbitrary coordinate system, yielding the coupling \eqref{eq:Hfull}.


\subsection{Master equation due to gravitational time dilation}
Here we derive an equation of motion for the centre-of-mass of a composite quantum system in the presence of time dilation. We keep the composition, the centre-of-mass Hamiltonian $H_{cm}$ and the relativistic time dilation completely general. The overall Hamiltonian of the system is $H_{tot}= H_{cm} +H_0 + H_{int}$, where $H_0$ governs the evolution of the internal constituents and $H_{int}= H_0\Gamma(x,p)/c^2$ captures the time-dilation-induced coupling between internal degrees of freedom and the centre-of-mass to lowest order in $c^{-2}$.
$\Gamma$ is a function of the centre-of-mass position $x$ and momentum $p$ to which the internal states couple due to special relativistic and general relativistic time dilation
(for the Schwarzschild metric in the weak-field limit we have  $\Gamma(x,p) = \Phi(x) - p^2/2m^2$). We start with the von Neumann equation for the full state  $\dot{\rho}= - i/\hbar \comm{H_{tot}}{\rho} $ and write  $H_{tot}= H + H_{int}$, where $H = H_{cm} +  H_0$. We change frame to primed coordinates, which we define through $\rho'(t) = e^{it(H +h)/\hbar} \rho(t) e^{-it(H +h)/\hbar}$, where $h =h(x,p)= \Pi_{i=1}^{N}\Tr_i{[H_{int} \rho_i(0)]} =\Gamma(x,p) \bar{E}_0 /c^2$ with the average internal energy $\bar{E}_0$. The resulting von-Neumann equation is
\be
\begin{split}
\dot{\rho}'(t) & = \frac{i}{\hbar} \comm{H'(t) + h'(t)}{\rho'(t)} -  \frac{i}{\hbar} \comm{H'(t) + H'_{int}(t)}{\rho'(t)} \\
& = - \frac{i}{\hbar} \comm{H'_{int}(t) - h'(t)}{\rho'(t)} \, ,
\end{split}
\ee{}
where $h'(t)=h(x'(t), p'(t))$. The formal solution $\rho'(t) =\rho'(0) - \frac{i}{\hbar} \int_0^t ds \comm{H'_{int}(s) - h'(s)}{\rho'(s)} $ is used in the  equation above, which yields the integro-differential equation
\be
\begin{split}
& \dot{\rho}'(t)   = - \frac{i}{\hbar} \comm{H'_{int}(t) - h'(t)}{\rho'(0)} \\
& - \frac{1}{\hbar^2} \int_0^t ds \comm{H'_{int}(t)- h'(t)}{\comm{H'_{int}(s) - h'(s)}{\rho'(s)}} .
\end{split}
\ee{}
We can now trace over the internal degrees of freedom. The state is initially uncorrelated $\rho(0) = \rho_{cm}(0) \otimes \Pi_i^N\rho_{i}(0)$ and we take the Born approximation, keeping only terms to second order in $H_{int}$. In this case $\rho'(s)$ can be replaced under the integral by $\rho'_{cm}(s) \otimes \rho_{i}(0)$ and the master equation for the centre-of-mass becomes
\be
\begin{split}
&  \dot{\rho}'_{cm}(t)  = \underset{i=1}{\overset{N}{\prod}} \Tr_i[\dot{\rho}'(t)]   \\
& \approx  \frac{-1}{\hbar^2} \underset{i=1}{\overset{N}{\prod}} \! \int_0^t \! \! \! ds \,  \Tr_i \Big\{\! \! \comm{H'_{int}(t) \!-\! h'(t)}{\comm{H'_{int}(s) \!-\! h'(s)}{\rho'(s)}} \! \! \Big\}  \\
&  = - \left(\frac{ 1}{\hbar c^2} \right)^2 \! \underset{i=1}{\overset{N}{\prod}} \! \int_0^t \! \! \! ds \, \Tr_i \Big\{ \! \! \left(H_0 \! - \! \bar{E}_0 \right)^2 \comm{\Gamma'(t)}{\comm{\Gamma'(s)}{\rho'(s)}} \! \! \Big\}  \\
&  = - \left(\frac{ \Delta E_0}{\hbar c^2} \right)^2  \int_0^t \! \! \! ds \,  \comm{\Gamma'(t)}{\comm{\Gamma'(s)}{\rho'_{cm}(s)}}  .
\end{split}
\ee{}
Here we used the notation $\Delta E_0^2 = \Pi_{i=1}^{N} \Tr_i \left\{ \left(H_0 - \bar{E}_0 \right)^2 \right\} = \mean{H_0^2} - \mean{H_0}^2$ for the energy fluctuations of the internal states and $\Gamma'(s) = \Gamma(x'(s), p'(s))$. Changing back to the Schr\"{o}dinger picture, and introducing $s \rightarrow t-s$ we obtain the integro-differential equation:
\be
\begin{split}
 & \dot{\rho}_{cm}(t)  = -\frac{i}{\hbar}\comm{H_{cm} + \Gamma(x,p) \frac{\bar{E}_0 }{ c^2}}{\rho_{cm}(t)}  \\
 & -  \left(\frac{ \Delta E_0}{\hbar c^2} \right)^2 \int_0^t \! \! ds  \comm{\Gamma(x,p)}{\comm{\Gamma(x,p)}{\rho_{cm}(t-s)} \! \big|_s} ,
\end{split}
\ee{eq:MEFull_app}
where $\comm{\Gamma}{\rho_{cm}} \! \big|_s = e^{- is H_{cm} /\hbar} \comm{\Gamma}{\rho_{cm}} e^{is H_{cm} /\hbar}$. This is the general equation of motion for a composite particle of arbitrary composition that undergoes time dilation. The decoherence of its off-diagonal elements is governed by its internal energy spread $\Delta E_0$ and by the metric-dependent coupling $\Gamma$.
The former is $\Delta E_0^2  \approx N (k_B T)^2$ in the high-temperature limit for $N/3$ non-interacting internal harmonic oscillators and the latter is $\Gamma = \Phi(x) \approx gx$ for stationary particles in the homogeneous weak-field limit of the Schwarzschild metric.


\subsection{Effect of time dilation on clocks and thermal states}
The time-dilation-induced coupling between internal degrees of freedom and the centre of mass causes decoherence of the latter. The complementarity between the visibility $V$ of interference and the which-path information $D$ is given by the inequality $V^2+D^2\leq 1$ \cite{ref:Englert1996}. The equal sign holds for pure states, i.e. for well-defined clocks as considered in ref.~\cite{ref:Zych2011} in the main text. For mixed states, it is possible to have loss of visibility with no accessible which-path information, as is the case here (as well as in most other decoherence models for which the Born approximation is used). The thermal state of the internal degrees of freedom can be seen as a mixture of clock-states each measuring the proper time along its respective path. To highlight this, consider for example a particle with a single 2-level internal degree of freedom which is in a clock-state, i.e. in a superposition of the ground and excited state with transition frequency $\omega$ and an arbitrary relative phase $\phi$:  $\ket{E_{\phi}} =\frac{1}{\sqrt{2}} \left(\ket{g} + e^{i \phi} \ket{e}\right) $. If the particle moves along a world line with overall proper time $\tau$, the internal state will evolve to $\ket{E_{\phi}(\tau)} =\frac{1}{\sqrt{2}} \left(\ket{g} + e^{i (\phi+ \omega \tau)} \ket{e}\right) $. For the particle in superposition along two paths with proper time difference $\Delta \tau$, the internal clock state will therefore acquire which-path information, thus leading to a loss in visibility given by $|\braket{E_{\phi}(\tau_1)}{E_{\phi}(\tau_2)}|=|\cos(\omega \Delta \tau /2 )|$, independent of the phase $\phi$.
For a fully mixed internal state (analogous to a thermal state),
\be
\rho = \frac{1}{2}\left( \ket{E_{\phi}}\bra{E_{\phi}} +\ket{E_{\phi+\pi}}\bra{E_{\phi+\pi}}\right),
\ee{eq:mixed}
the relative phase between $\ket{g}$ and $\ket{e}$ is unknown and thus no which-path information is available, but it still results in a drop in visibility. The above state can be equivalently written in the basis
\be
\rho = \frac{1}{2}\left( \ket{g}\bra{g}+\ket{e}\bra{e}\right).
\ee{eq:mixed2}
Since \eqref{eq:mixed} and \eqref{eq:mixed2} represent the same state, they cannot be discriminated and will result in the same loss of visibility for the centre of mass. The states $\ket{g}$ and $\ket{e}$ individually, however, are not clock-states, thus no time dilation can be read out directly. The interpretation of the visibility drop in this representation is the phase scrambling due to the red shift, since a system with internal states $\ket{e}$ has different weight than if the states were $\ket{g}$ and thus acquires a different phase shift.  Irrespective of the state representation, time dilation couples the internal states to the centre of mass and thus cause decoherence of composite particles with internal degrees of freedom.

To highlight that time dilation can physically affect thermal states, we re-derive in our formalism the well known classical, general relativistic and thermodynamical effect first considered by Tolman  \cite{ref:Tolman1930}.
By definition, the dynamics of internal degrees of freedom is governed by the Hamiltonian $H_0$ in the local rest-frame of the system. If the system is thermal, its state is therefore $\rho_0 \propto e^{-H_0/k_B T_0}$, where the subscript $0$ highlights that the quantities are with respect to the local rest-frame (which is the temperature used in the main text). However, from the point of view of a stationary observer (laboratory frame) at some distance, the internal dynamics of the system at height $x$ in a gravitational potential $\Phi(x)$ is governed by the Hamiltonian $H_0 \sqrt{-g_{00}} \approx H_0(1+\Phi(x)/c^2)$. Thus the state in that frame is given by $\rho_{\text{lab}}(x) \propto e^{-H_0(1+\Phi(x)/c^2)/k_B T_{\text{lab}}}$. But since the density matrix assigned to a system does not depend on the reference frame (it can be interpreted in terms of occupation numbers of its eigenstates, irrespective of the energy labels assigned to them), we have $\rho_{0}(x) = \rho_{\text{lab}}(x)$, from which it follows that local and laboratory temperatures are related by $T_{\text{lab}}=T_{\text{0}}(1+\Phi(x)/c^2)$. The condition of thermal equilibrium for systems with different Hamiltonians is simply that they have the same temperature, thus two systems at heights $x_1$, $x_2$ are in thermal equilibrium for an outside observer if $T_{\text{lab}}(x_1) = T_{\text{lab}}(x_2)$. Hence, in terms of the locally measured temperatures, the condition for thermal equilibrium between the two systems is $T_{\text{0}}(x_1)(1+\Phi(x_1)/c^2)=T_{\text{0}}(x_2)(1+\Phi(x_2)/c^2)$, which means that $T_{\text{0}}(x_1)$ and $T_{\text{0}}(x_2)$ are different for systems in different gravitational potentials. For a single extended system in thermal equilibrium in a gravitational field the locally measured temperatures satisfy $T_{\text{0}}(x)(1+\Phi(x)/c^2)=$ constant, which is the Tolman law.

This classical effect can be seen as caused by the ``weight of heat'' or gravitational redshift  -- as in Tolman's own work  \cite{ref:Tolman1930}  -- or, equivalently, as caused by time dilation between parts of the  system at different gravitational potentials: the local internal frequencies $\omega_0$  are time dilated with respect to the laboratory reference frame, in which they become $\omega_0(1+\Phi(x)/c^2)$.  For the system to be in equilibrium the radiation coming from any part of the system must have the same frequency spectrum. The local quantities $\omega_0$ must therefore differ with height and satisfy $\omega_0(x)(1+\Phi(x)/c^2)=$ constant.
The Tolman effect and the decoherence effect discussed in this work have the same origin: time dilation affecting thermal states in the presence gravity.

\end{document}